\documentstyle[aps,multicol,epsf,eqsecnum,prl]{revtex}

\begin{document}

\renewcommand{\aa}{{\bf a}}
\renewcommand{\ss}{{\bf s}}
\def \LS {{\bf L}}     
\newcommand{\LSbase}{{\LS^{\rm base}}}
\newcommand{\LSprime}{{\LS'}}
\renewcommand{\lambdabar}{{\overline{\lambda}}}
\def \BB {{\bf B}}
\def \MM {{\bf M}}
\def \QQ {{\bf Q}}
\def \DC {{\cal D}}
\def \JC {{\cal J}}
\def \HalfO {{\cal O}}
\def \Isolated{{\cal O}'}

\def \bee {\begin{equation}}
\def \eee {\end{equation}}
\def \bea {\begin{eqnarray}}
\def \eea {\end{eqnarray}}
\newcommand{\qq}{{\bf q}}
\newcommand{\rr}{{\bf r}}
\def \mm {{\bf m}}
\def \nn {{\bf n}}
\def \XX {{\bf X}}
\def \tXX {{\tilde \XX}}
\def \Dgc {{D_{gc}}}
\def \tXXg {{\tXX_g}}
\def \YY {{\bf Y}}
\def \nusite{{\nu_{\rm site}}}
\def \Ksim{{K_{\rm sim}}}
\def \ptri {p_{\rm tri}}
\def \pstar {p_*}
\def \yy {{\bf y}}
\def \Fcal {{\cal F}}
\def \Gcal {{\cal G}}
\def \SC   {{\cal S}}
\def \HH {{\cal H}}
\newcommand {\mueff}{\mu_{\rm eff}}
\newcommand {\HHbiq}{\HH_{\rm biq}}
\newcommand {\HHdil}{\HH_{\rm dil}}
\newcommand {\HHquant}{\HH_{\rm quant}}

\newcommand{\jc}{j}    
\newcommand{\SS}{{\bf S}}
\newcommand{\Stot}{{\SS_{\rm tot}}}
\newcommand{\Xdef}{x_{\rm def}}
\newcommand{\Xdefiso}{x_{\rm def}'}

\title{Effective Hamiltonians and dilution effects in 
kagom\'e and related antiferromagnets}

\author{Christopher L. Henley}
\address{Laboratory of Atomic and Solid State Physics, Cornell University,
Ithaca, New York, 14853-2501}

\date{\today}

\maketitle

\begin{abstract}
What is the zero-temperature ordering pattern of
a Heisenberg antiferromagnet with large spin length
$S$ (and possibly small dilution), on the kagom\'e lattice, 
or others built from corner-sharing triangles and tetrahedra?
First, I summarize the  uses of effective Hamiltonians 
to resolve the large ground-state degeneracy,
leading to long-range order of the usual kind.
Secondly, I discuss the effects of dilution,
in particular to {\it non}-frustration of classical ground states,
in that every simplex of spins is optimally satisfied. 
Of three explanations for this, 
the most complete is
Moessner-Chalker constraint-counting.
Quantum zero-point energy may compete with 
classical exchange energy in a diluted system, 
creating frustration and enabling a spin-glass state. 
I suggest that the regime of over 97\% occupation is
qualitatively different from the more strongly diluted regime.
\end{abstract}

\begin{multicols}{2}

\section{Introduction}

This paper addresses the isotropic 
antiferromagnet with quantum Heisenberg spins $\SS_i$, 
on kagom\'e  and analogous lattices (``bisimplex'' 
lattices, to be defined shortly.) 
Everything is restricted to the large-$S$ and 
low-$T$ limit, where $S$ is the spin quantum number and 
$T$ is the temperature.  
Thus, to lowest order we may visualize
$\SS_i \approx S \ss_i$, 
where $\ss_i$ is a classical vector of unit length.

The theme of Sec.~\ref{sec-Heff} is the usefulness of 
{\it effective Hamiltonians}, in which some degrees of 
freedom are eliminated, in favor of new terms involving 
the remaining degrees of freedom.  
The reader is also reminded that the classical picture can be 
qualitatively wrong, at the temperatures of interest, which are well below 
the quantum spinwave energies.
For example, the quantum pyrochlore lattice is seen to be {\it less}
degenerate than the kagom\'e case, in contrast to classical results;
and the effective interactions favoring collinear/coplanar and 
other forms of order are much more powerful in the quantum case.

The rest of the paper concerns the effects of dilution.
Does it produce a spin glass, or generate an 
effective Hamiltonian favoring long-range order, or 
preserve the exceptional degeneracy of the pure lattice?
Sec.~\ref{sec-DilOrd} reviews the effect of dilution in ordinary
frustrated systems, to contrast with its effect 
in bisimplex lattices (Sec.~\ref{sec-DilBisimp}), 
in which the simplex units are all satisfied.
Sec.~\ref{sec-WhySat}, 
the heart of this paper, 
presents three explanations of the simplex satisfaction.
But when we admit the full zoo of real effects --
quantum fluctuations, dilution, 
unequal exchange constants, anisotropies, external field --
it is likely that the simplices stop being satisfied
(Sec.~\ref{sec-frust}).

``Lattices analogous to the kagom\'e'' meant, more
precisely,  the {\it bisimplex}
lattices: those derived from a bipartite network
by placing a spin at each bond midpoint, so that
each spin belongs to two {\it simplices}.
(My ``simplex'' is a synonym for ``unit'' as
used by Moessner {\it et al}~\cite{moess98,moessber99}.)
The coordination number in the network becomes
the number of corners $q$  of each {\it simplex}, which means
a single  bond, a triangle, or a tetrahedron
for $q=2$, $3$, or $4$, respectively.
Table \ref{t-bisimplex} lists bisimplex lattices
mentioned in the literature (there are more), tagged
by the dimensionality and the $q,q'$ values for the 
simplices on either side of a site.

The kagom\'e  and pyrochlore lattices are familiar;
the connected magnetic sites in $\rm SrCr_{9p}Ga_{12-9p}O_{19}$ (SCGO) 
form the $d=2$ bisimplex lattice that I'll
call ``kagom\'e  sandwich'' (also known as ``pyrochlore slab'').
It consists of two kagom\'e  layers connected by a 
a triangular-lattice linking layer~\cite{shlee96}. 
The three-dimensional magnetic lattice of
{\it e.g.} gallium gadolinium garnet~\cite{GGG}  is
appropriately dubbed ``hyperkagom\'e''~\cite{hyperkag}, 
since it too consists of corner-sharing triangles.
The crossed-square lattice 
is a pyrochlore slab normal to $\{ 100\}$,  with 
its top and bottom surfaces identified, and serves as a
two-dimensional toy model for the pyrochlore.
(See Fig.~2 of Ref.~\onlinecite{moess98}(b))

The site-percolation threshold $p_c$ of the bisimplex lattice
is obviously the bond-percolation $p_c$ of its parent bipartite network.
It is listed in Table~\ref{t-bisimplex} because
one expects qualitative changes of behavior at $p_c$, if the spin system 
orders in any fashion.~\cite{keren00}
The sandwich lattice $p_c$ is published here
for the first time, to my knowledge. (See Appendix~\ref{app-pc}.)

The Hamiltonian couples nearest neighbors,
   \bee
      \HH = j \sum _{\langle ij \rangle} \SS_i \cdot \SS_j 
   \eee
or classically,
with a magnetic field $\BB$ included, 
   \bee
      \HH = J \sum _{\langle ij \rangle} 
            \ss_i \cdot \ss_j -\BB \cdot \sum _i\ss_i 
           = \sum _\alpha \frac{J}{2} |\LS_\alpha-\frac{\lambda}{J}\BB|^2 
              +E_0
       \label{eq-HHLsq}
   \eee
where the exchange constant is $J\equiv jS^2 > 0$, 
$\lambda = 1/2$, and the total spin of simplex $\alpha$ is
   \bee
              \LS_\alpha \equiv \sum _{i \in \alpha} \ss_i .
   \label{eq-LS}
   \eee
In the kagom\'e-sandwich case, 
the interlayer (kagom\'e to linking layer) 
coupling will be called $J'$, but unless explicitly noted, 
I assume $J'=J$, the kagom\'e-layer coupling. 
    
Frustration means generally that, having broken up the Hamiltonian 
into local terms, we cannot simultaneously satisfy all of them.
In the present case, each term in 
(\ref{eq-HHLsq}) is satisfied (in zero field) whenever $\LS_\alpha=0$.
Then each undiluted bisimplex lattice is completely {\it unfrustrated} 
from the simplex viewpoint, since it can be shown (by example)
that every simplex {\it can} attain this minimum at the same time.
Let $\XX$ denote the
classical ground state manifold (for $\BB=0$); 
it is massively degenerate on every bisimplex lattice.

\section {Effective Hamiltonians and order by disorder}
\label{sec-Heff}

Our goal is to discover how the system breaks symmetry and 
orders (if it does). 
We assume provisionally that, for sufficiently large $S$, the
quantum ground state is one of the classical ground states, dressed by
some quantum zero-point spin fluctuations. 
(This clearly fails if the fluctuations about a classical ground state
are as big as the distance to the next one, or if 
the system tunnels so freely between dressed classical states 
that the correct wavefunction is a superposition.)
This assumption directly gives us the ordering pattern
in a simple (e.g. triangular lattice) Heisenberg antiferromagnet, 
where the classical ground state is unique (modulo symmetries). 

But commonly, the classical ground states have nontrivial degeneracies, 
so that every ground state has a different manifold of possible 
small spin deviations  and consequently a different zero-point energy.  
Presumably, the true quantum ground state should be
constructed around the particular (ordered!) classical ground state
which has the lowest zero-point energy. 
When the thermodynamic state of the system has true long-range
order in a particular pattern, due to this fluctuational energy,
we could call it ``order by disorder''~\cite{vill80,hen89}.
(This usage of the term is broader in some ways, and narrower in others,
than the ``local'' definition in Ref.~\onlinecite{moess98}.)

To transparently model the degeneracy-breaking, one
may construct effective Hamiltonians in closed form
(via intelligently applied perturbation theory). 
They are intended, not to provide the accurate energy of a special state
or two, but as inputs to further modeling, 
{\it e.~g.} simulations at $T>0$, or tunneling calculations~\cite{vd93}. 

\subsection {Harmonic and higher order effective Hamiltonian(s)}

The spin-wave expansion naturally organizes the zero-point energy 
as an expansion in powers of the small parameter $1/S$, 
of which the zero term is the $O(\jc S^2)$ 
classical energy $E_0$ and
the first term is the $O(\jc S)$ {\it harmonic} zero-point energy
$\Fcal(\XX) = \frac{1}{2} \sum _k \hbar \omega_k$, 
including all spin-wave frequencies $\omega_k$ (which depend
implicitly on $\XX$.)
This $\Fcal(\XX)$ is the effective Hamiltonian, 
defined only on states in the 
ground state manifold.~\footnote{
Analogously,  in the classical model
with $T/J \ll 1$ 
one defines a harmonic free energy $\Fcal \sim T \ll E_0 \sim J$, 
by integrating out some degrees of freedom
in the partition function.  See Ref. ~\onlinecite{shenderHeff}.}
Call the {\it local} minima of $\Fcal()$ the ``favored states'' 
$\{ \YY \}$; these are a discrete set, 
(in all cases I know),
modulo global rotation symmetry.

Let ``ordinary degenerate antiferromagnet''  mean one
for which $\{ \XX \}$ is a finite-dimensional manifold
-- clearly {\it every} ground state is periodic.
Examples are the FCC antiferromagnets~\cite{hen87,larson90}  
or the two-sublattice systems with 
second-neighbor interactions~\cite{hen89,shender82,yjkim}.
(A ``highly frustrated'' system might be defined as one
in which the dimensionality of $\XX$ is extensive, 
and two ground states may differ only locally.) 
For an ordinary degenerate system, with isotropic
exchange interactions,  a crude approximation for
$\Fcal(\XX)$ is the {\it biquadratic effective Hamiltonian}
   \bee
      \HHbiq \equiv - \sum _{ij} K_{ij} (\ss_i\cdot \ss_j)^2, 
   \label{eq-Hbiq}
   \eee
This was independently posited phenomenologically~\cite{jacobs};
it can be obtained analytically in a couple of
ways using a perturbation expansion~\cite{larson90,zhang00}
around mean-field theory. 
One obtains $K_{ij} = S^2 j_{ij}^2/8 h_{\rm loc}$, 
where the local field $h_{\rm loc}$ is $2 jS$ in a bisimplex lattice.
Eq.~(\ref{eq-Hbiq}) correctly tells us that the 
favored states are the {\it collinear} ones.
In the  kagom\'e case.
(\ref{eq-Hbiq}) should be replaced by a
different functional form to approximately represent $\Fcal(\XX)$, 
because (i) the criterion for favored states is that the
spins are {\it coplanar}; 
(ii) the true function
$\Fcal (\YY+\delta\XX)-\Fcal(\YY) \sim |\XX-\YY|$, linear~\cite{vd93,Ri93}
rather than quadratic in the deviations $\delta \XX$. 

More fundamentally, the harmonic approach 
fails in the kagom\'e case because it does not
fully break the degeneracy. 
Every coplanar state has exactly the same harmonic-order 
Hamiltonian, if written using as coordinates at each site
(i) the $y$ (out-of-plane) spin deviation
and (ii) the rotation angle around the $y$ axis.
The source of this,  mathematically, is that 
$\ss_i\cdot \ss_j$ takes the same value (here $-1/2$) 
for every nearest-neighbor pair, in every favored ground state.
Consquently, on a bisimplex lattice with only triangles, 
{\it i.e.,} the kagom\'e or hyperkagom\'e lattices, 
$\Fcal({\YY})$ has exactly the same value for {\it every} 
coplanar state $\YY$.
The number of such states is $O(\exp ~N)$, where $N$ is
the number of unit cells.

To resolve the surviving degeneracy among favored 
ground states, we need a second effective Hamiltonian
$\Gcal (\YY)$, obtained from some sort of self-consistent theory
that takes account of anharmonic spin-wave interactions
~\cite{chubukov,chan,henchan94ICM}.
Since $\{ \YY \}$ is discrete, $\Gcal(\YY)$ is parametrized by 
discrete-spin variables: 
in the kagom\'e case,
the ``chiralities'' $\tau_\alpha = \pm 1$
defined from the spins on each plaquette~\cite{Ri93}.
Our own approximation~\cite{chan,henchan94ICM} gave
   \bee
     \Gcal(\{\tau _\alpha\}) = -\sum _{\alpha\beta} 
     \JC(R_{\alpha\beta}) \tau_\alpha \tau_\beta,
   \eee
which has the form of an antiferromagnetic Ising Hamiltonian on 
the honeycomb lattice, 
and is defined for {\it every} favored ground state $\YY$. 
The energy scale~\cite{chubukov} is certainly $\JC = O(jS^{2/3})$, i.e. down 
by only a factor $S^{-1/3}$ from the scale of the 
harmonic term $\Fcal(\XX)$.

In the pyrochlore case, Eq.~(\ref{eq-Hbiq}) is also too crude:
here the favored ground states $\YY$ {\it are} the collinear ones, 
but their harmonic energies $\Fcal(\YY)$ are nondegenerate since
different collinear states have different 
patterns of $\ss_i\cdot \ss_j = \pm 1$.
It turns out that
the special states that minimize $\Fcal()$ are still infinite in number, 
but only as $\exp({\rm const} ~L)$, 
where $L$ is the system's diameter~\cite{Hen00pyharm};
the kagom\'e-sandwich lattice seems to behave similarly.
Thus, in the quantum case, the pyrochlore and sandwich lattices 
show {\it more} ordering tendency than the kagom\'e lattice 
(at harmonic order), whereas in the classical case
it is the other way around~\cite{moess98}. 

\subsection{Pitfalls of classical modeling}
\label{sec-pitfalls}

Although large $S$ justifies visualizing each spin as a
fixed-length vector, it does {\it not} justify 
a purely classical simulation of the system. 
The reason is that the interesting phenomena occur 
when $T \ll jS$, the scale of $\Fcal(\XX)$. 
Thus thermal effects are only a small correction to the quantum effects.
I believe there is an easy fix: 
a classical Monte Carlo simulation using a Hamiltonian which 
includes (an approximation of) $\Fcal(\XX)$ and $\Gcal(\YY)$
ought to give valid physical results.

For a specific example, consider SCGO with $S=3/2$ and
a Curie-Weiss constant~\cite{ramirez90} of $515{\rm K}$, 
hence $j \approx 80{\rm K}$. 
Then,  in a kagom\'e lattice, the effective energy of
coplanarity would be $0.14 jS$ per spin along 
a straight ``spin fold''~\cite{Ri93}; if we presume this
also applies to the shortest rotatable loop of six sites, we get
a barrier of $100 {\rm K}$ (alternatively $\frac{5}{2}jS \approx 300{\rm K}$
from App. B of Ref.~\onlinecite{vd93}). 
That is vastly larger than the
analogous free energy barrier in a classical system, 
at the SCGO freezing temperature $\sim 3.5 {\rm K}$.  
Again, in a pyrochlore lattice with the same coupling $j$, 
the coefficient in (\ref{eq-Hbiq}) comes out to 
$K_{ij} = \frac{1}{16}jS \approx 7.5 {\rm K}$ per nearest-neighbor bond.
That ought to induce a transition to long-range-ordered collinearity
at a temperature of order $10 {\rm K})$, which wouldn't happen
at any temperature in a classical system~\cite{moess98}. 

\section{Dilution in ordinary cases}
\label{sec-DilOrd}

I now turn to the longer part of this paper: what 
effect(s) does dilution have on a highly frustrated system, 
specifically on a bisimplex-lattice antiferromagnet?
Can it be represented by an effective Hamiltonian?
Or can it turn an ordered system into a spin glass?
The answers are ``yes'' for the ``ordinary'' 
antiferromagnets, which are reviewed in this
section (in three regimes). 
The answers are different for highly frustrated magnets
(later sections).

Starting with this section, I am considering
{\it classical} ground states at $T=0$
(unless explicitly noted).

\subsection{Unfrustrated case}

In an {\it unfrustrated} 2-sublattice (N\'eel) antiferromagnet
(Fig.~\ref{fig-ordinary}(a)), 
the ground state at occupied fraction $p$ 
(on the unique extended cluster) 
has exactly the same spin directions as at $p=1$, 
on all the  magnetic sites. 
The total magnetization
does not cancel exactly, since the moment
on the even or odd sublattice has $O(\sqrt{N})$ 
statistical fluctuations.
An observable corollary of this observation is that the structure factor
at wavevector $\qq$, 
   \bee 
       \SC(\qq) \equiv 
            \frac {1}{N}
             \langle |\sum _i e^{i\qq\cdot{\rr_i}} 
                \ss_i |^2 \rangle, 
   \eee
has the limiting behavior
   \bee
       \lim _{\qq \to 0} \SC(\qq) = p(1-p) > 0.
   \label{eq-Sq0}
   \eee

\subsection{Frustrated case: weak dilution}
\label{sec-ordweak}

Consider next a simple frustrated case, {\it e.g.}
a Heisenberg triangular antiferromagnet.
The pure system has a periodic, non-collinear ground state in which the 
spins differ by $120^\circ$ angles, which is the best compromise
within each triangle of spins.
The spins in the diluted system no longer take the same
directions they would in the pure system.  

One regime is weak dilution ($p$ close to 1). Consider
just one nonmagnetic site (Fig.~\ref{fig-ordinary}(b)).
Far away from this defect, the configuration essentially
agrees with a pure ground state.
But the neighbors of the removed spin deviate towards the direction
it would have had, and their neighbors deviate in turn.
Thus the defect creates a slowly-varying spin twist
with a pseudo-dipolar spatial dependence.
(It has the angular dependence of a dipole field and
decays with distance as $1/|\rr|^d$, where $d$ is the spatial dimension.)

In the ordinary degenerate antiferromagnets, 
the energy reduction due to spin deviations 
depends on which ground state one deviates from.
The average of this energy,  over all the ways to 
place a low density $\Xdef$ of defects, 
defintes an {\it effective Hamiltonian}
function $\HHdil(\XX)$, proportional to $\Xdef$
and possessing the full symmetry of the ground states
$\XX$ of the {\it pure} system.
In exchange-coupled systems, 
$\HHdil(\XX)$ favors the least collinear ground states, 
and thus has an interesting competition with $\HHquant$~\cite{hen89,larson90}.

Often, depending on how a defect's local 
symmetry relates to that of the ground-state manifold $\XX$, 
each defect may prefer
a particular ground state, out of the subspace $\{ \YY \}$ favored
by $\HHdil$.
In that case, dilution creates an
``effective random field''~\cite{hen89,larson90}
varying in space and coupling to the discrete degrees
of freedom $\YY$.
As in other random-field models, disorder wins
if it is sufficiently strong or the dimension is low enough.
The resulting state is a spin glass, in the sense that
it lacks long-range correlations and has barriers, but
has not been proven to be the same phase as a 
$\pm J$ spin glass~\cite{wengel96}.

\subsection{Frustrated case: strong dilution}
\label{sec:ord-strong}

Now return to the same triangular lattice, but with strong dilution
so that $p$ is just above $p_c$. 
It is well known that,  near percolation, 
the typical connection between two sites (if any) is tenuous, 
and order is propagated over one-dimensional chains of sites, 
which are multiply-connected at occasional places.
At $T=0$, the spin directions alternate along such a chain, so
it constrains the relative orientation of the endpoint spins to
be parallel or antiparallel, depending
on whether the number of bonds connecting them is even or odd.
Order is propagated (at $T=0$) as if there were 
a direct bond between the endpoints.
But if there are {\it two} paths in parallel, 
they may disagree on the relation between endpoint spins;
the smallest example is shown in Fig.~\ref{fig-ordinary}(c).
The ground state is a twisted spin-configuration that is found
neither in the pure lattice nor in a singly-connected chain. 
In Fig.~\ref{fig-ordinary}(c), 
the effect is to force an $80^\circ$ angle between the endpoint spin directions.

The extended connected cluster is an irregular network containing loops
within loops of this type.  It is plausible 
that its global ``energy landscape'' is like that of a spin glass, 
possessing numerous low-lying, 
nearly degenerate energy minima, separated by energy barriers. 

\section{Dilution in a bisimplex lattice}
\label{sec-DilBisimp}

Dilution affects a bisimplex lattice quite differently 
than an ordinary frustrated lattice:
simulations find that essentially {\it every simplex remains satisfied}, 
i.e. $\LS_\alpha=0$, even for strong dilution.
(This includes the simplices with nonmagnetic sites:
they are still simplices, but the number of corners $q$ is reduced.)
In this section, I will discuss the evidence for, and some
corollaries of, that fact.

\subsection {Simulations of 2-layer kagom\'e lattice}
\label{sec-simulations}

The original evidence was that (in the kagom\'e lattice)
the local field is exactly $2$ on most sites ~\cite{huber93,Sh93}.
I carried out more extensive 
simulations like Ref.~\onlinecite{huber93}, 
for the diluted kagom\'e-sandwich lattice
-- relevant to experiments on SCGO.
(High-quality crystals of that material
can be grown only with $p<1$.) 
Over 500 independent realizations of the disorder were constructed, 
for a $10\times 10$ lattice with $p=0.55$
(i.e., 385 spins);
each realization was relaxed 
from three different random initial configurations.
to a ground state by $250$ sweeps, in which 
each site in turn was set to its local-field direction.
The program flags all configurations in 
which $|\LS_\alpha| > 0.1 $ on any simplex with $q>1$.
This happened only on ``one-eared'' loops like Fig.~\ref{fig-earloops}(a).
\footnote{Three exceptions are mentioned at the end of 
Sec.~\ref{sec-constrprop}.}
Now, the ``one-eared'' loop connects to the rest
of the world in (at most) one point. 
Hence it can't induce twisted, frustrated relationships 
among distant spins (like Fig.~\ref{fig-ordinary}(c)), 
and can't be responsible for spin-glass behavior.
Similar results were found even when $J'\neq J$ (unequal 
interlayer and kagom\'e layer exchange), as well as
in a plain kagom\'e lattice.

\subsection {``Half-orphan spins''}
\label{sec-orphan}

Simplex satisfaction has observable consequencies. 
Every simplex has $\LS_\alpha=0$ -- except, of course, 
that a $q=1$ (one-spin) simplex has $|\LS_\alpha|=1$. 
Let $\HalfO$ be the set of spins which have $q=1$ on one side, 
with frequency $\Xdef \sim (1-p)^2$ per unit cell.
Also let $\Isolated$ be the spins which are completely isolated
($q=1$ on both sides), with frequency $\Xdefiso \sim (1-p)^5$
in the sandwich lattice -- i.e., rare for $p> p_c$.
The total magnetization (in units of $\mu \equiv (2 \mu_B) S$) is
   \bee
      \MM_{\rm tot} = \frac{1}{2} \sum _\alpha \LS_\alpha =
       \frac{1}{2} \sum_ {i\in \HalfO}  \ss_i +
       \sum_ {i\in \Isolated}  \ss_i 
   \label{eq-Morphan}
   \eee
The prefactors of $1/2$ in (\ref{eq-Morphan}) appear because
each spin's moment is divided between two simplices.
Schiffer and Daruka~\cite{sch97orphan}
observed a Curie law contribution to the susceptibility, 
ascribed to ``orphan spins'' by them (which suggests
the isolated spins), but more plausibly to the spins in
$\HalfO$~\cite{moessber99,mendels}, which 
might better be called ``half-orphan'', since 
they belong to a simplex on one side but are 
isolated on the other side. 
If these moments had independent directions
(but see Sec.~\ref{sec-gauss}), 
they would produce the Curie susceptibility, per cell, 
   \bee
         \chi(T) = \frac {\mueff^2}{3T} \left(\frac{1}{4}\Xdef+\Xdefiso\right) .
   \label {eq-naiveCurie}
   \eee

Insofar as the system is built from satisfied simplices, its
total (classical) magnetization is zero. 
So, in place of (\ref{eq-Sq0}), 
the structure factor (\ref{eq-Sq0}) scales
as $\Xdef \ll (1-p)$. 
Thus even with significant dilution,  one expects
$\SC(\qq \to 0) \approx 0$, 
as seen in SCGO
(according to Ref.~\onlinecite{shlee96}). 

The half-orphan site has an free spin of $S/2$ in
(\ref{eq-Morphan}) and (\ref{eq-naiveCurie}).
This is a perfect classical analog of the spin-1/2 moment
that appears in diluted $S=1$ quantum spin chains with a valence-bond state
~\cite{vbsdefect}.
I speculate that quantum $S=1$ bisimplex lattices indeed have
a valence-bond state, and $S=1/2$ defect moments.

The simplex-satisfaction concept also encompasses bond-randomness effects
(see start of Sec.~\ref{sec-frust}). The ground state of a 
tetrahedron with one ferromagnetic bond still has $\LS_\alpha=0$
so it does not affect the near-cancellation of $\MM_{\rm tot}$
or the Gauss's law (Sec.~\ref{sec-gauss}).
On the other hand, a triangle with one ferromagnetic bond is
unfrustrated, with a configuration like
$\uparrow\uparrow\downarrow$.
Then $|\LS_\alpha|=1$: 
a ferromagnetic bond produces exactly the same sort of
paramagnetic defect as a ``half-orphan'' spin.

\subsection{Satisfaction with unequal couplings}
\label{sec-Jprime}

What happens to the picture of Eq.~(\ref{eq-HHLsq})
when $J'\neq J$ (coupling to linking layer 
and/or  field $\BB$ is nonzero?
It turns out (\ref{eq-HHLsq}) still works, provided we 
now take~\cite{moessber99}
    \bee
           \LS_\alpha \equiv \sum _{i\in \alpha} w_{\alpha i} \ss_i
    \eee
where $w_{\alpha i}= J'/J$ when $\alpha$ is (before dilution)
a tetrahedron and $i$ is the linking-layer spin that caps it, and 
$w_{\alpha i}=1$ otherwise.
Also, we must replace $\lambda$ in (\ref{eq-HHLsq}) by
$\lambda(\alpha) = J/2J'$ when $\alpha$ is (before dilution) a triangle, 
and $1-J/2J'$ otherwise (when $\alpha$ is, before dilution, a tetrahedron).
To satisfy each term, $\LS_\alpha = \lambda(\alpha)\BB$, 
which immediately implies (in the pure system)
    \bee
           \langle \ss_i \rangle = \cases 
                                    { \frac{J}{6J'} \BB , & 
                                          $i$ in kagom\'e layer;\cr
                                    \left(1-\frac{J}{J'}\right) \BB , & 
                                          $i$ in linking layer.\cr}
    \label{eq-layermeans}
    \eee
Notice that the sum of all linking-layer spins must be exactly zero either
if $\BB=0$ {\it or} if $J'=J$ (even in nonzero field). 
Also, the total magnetization of satisfied simplices
is (\ref{eq-Morphan}), except the coefficients $1/2$ must be replaced
by $\lambda(\alpha)$.
Thus, with $J'\neq J$ the net magnetization of the satisfied simplices 
{\it still} must be zero.~\footnote{
This is contrary to the conclusion of Ref.~\onlinecite{moessber99}.
Their calculation treated the simplices as independent.
That approximation can violate important sum rules,  e.g. 
it finds a mean spin on the tetrahedron-base which is different from
that on a triangle, even though these are in fact the same spins.}

There is a problem 
whenever $i$ is a linking-layer spin and $\alpha$  (after dilution) is
a $q=2$ simplex:
one must set $w_{\alpha i}=1$ again to correctly describe the satisfied bond.
But then, in a magnetic field,  
the decomposition into terms (\ref{eq-HHLsq}) breaks down since
there is no consistent value for $\lambda(\alpha)$.
Even for $\BB=0$, if $w_{\beta i}=J'/J$ still for the
other simplex that includes site $i$, the result (\ref{eq-Qthm}) 
of the next section breaks down; in effect, this site enters
(\ref{eq-Qthm}) as another sort of ``point charge''.

\subsection {Divergence theorem}
\label{sec-gauss}

I now introduce
a sort of ``Gauss's Law'', which is handy for revealing the
nonlocal effects of defects.
Recall that in a ``bisimplex lattice'', every spin
belongs to one even and one odd simplex;
let $(-1)^\beta \equiv  +1$ or $-1$ when $\beta$ labels an 
even or an odd simplex, respectively.
Mark out a domain $\DC$ containing a subset of the simplices.
Define a kind of ``charge'', 
\bee
    \QQ(\DC) \equiv \sum _{\beta \in \DC} (-1)^\beta \LS_\beta.
\label{eq-Qsum}
\eee
The theorem states that, assuming simplex satisfaction,
  \bee
      \sum _ {i \in \HalfO} (-1)^{\beta(i)} \ss_i = \QQ(\DC) = 
      \sum _ {i \in \partial \DC} (-1)^{\delta(i)} \ss_i, 
      \label{eq-Qthm}
  \eee
where $\partial \DC$ is the set of sites the domain boundary 
cuts through. Also, $\beta(i)$ tells which simplex has $q=1$ 
(of the two containing the half-orphan site $i$), 
and $\delta(i)$ tells which simplex is in the interior 
(of the two containing boundary site $i$). 
The left-hand side of (\ref{eq-Qthm}) follows since 
only the $q=1$ simplices contribute nonzero terms in (\ref{eq-Qsum});
the right-hand side follows because
every spin in the {\it interior} of $\DC$ appears in 
two terms of (\ref{eq-Qsum}) with canceling coefficients.
(Thus, half-orphan spins are the point charges in our ``Gauss's law'', 
while $(-1)^{\delta(i)} \ss_i$ plays the role
of the normal component of the electric field at the surface.)
This is a generalization of the sum rules of Ref.~\onlinecite{moess98}(b), 
Sec.~III B.\footnote{
Note also that, in the pure system (no ``charges''), 
a vector potential for the ``electric field'' can be constructed, 
which is uniquely valued if $d=2$;
in the kagom\'e case, this is just the ``spin origami''
embedding of Ref.~\onlinecite{Sh93}.}
By drawing a succession of nested boundaries $\nabla\DC$ around
a single half-orphan spin $\ss_I$, one shows that
neighboring spins $\ss_j$ 
have correlations with $\ss_I$ that alternate in sign 
(as speculated by Mendels~\cite{mendels})
and decay  with distance as $1/|\rr_j-\rr_I|^{d-1}$.

If we let $\DC$ include the entire system, 
there is no boundary term and the total ``charge'' 
(left-hand side of (\ref{eq-Qthm})) must be zero.
This implies that, if there is just one half-orphan spin, it is impossible
to exactly satisfy every simplex; if there are just two
half-orphan spins, they must be exactly parallel or antiparallel
(depending on the relative parity of their  respective $q=1$ simplices).
But this law has a very large loophole: the violation of $\LS_\alpha=0$
may be spread out uniformly over the simplices, such that the 
{\it total} energy cost is $O(1/N)$, which is negligible in a large
system. 
There is a more physical argument why nearby ``charges'' ought, 
nevertheless, to cancel, as the electrostatic analogy would suggest. 
If they don't cancel, 
spins in shells surrounding these ``charges'' are constrained by
(\ref{eq-Qthm}) to have a nonzero mean ``charge'', which would reduce the 
number of possible states,  thus reduce the {\it entropy}, thus
increase the {\it free energy} at $T>0$. 
I conjecture that, in the diluted {\it quantum} system,  there is 
an analogous effective interaction between nearby half-orphan spins, 
mediated by the harmonic zero-point energy.

\subsection{NMR experiments}

An NMR experiment measures the distribution of the local fields $h$
felt by each NMR nucleus (Ga, in SCGO). 
This is an average of the magnetizations of its neighbor (Cr) spins, 
$\mm_i = \langle \ss_i \rangle$, 
where the average is taken over all ground states (for a fixed
realization).
the variance of $h$ would scale as $\Xdef$, 
so the NMR linewidth should scale as 
$\sqrt{\Xdef} \sim 1-p$, as is seen experimentally~\cite{mendels}.

The experiments can separate the NMR signal from the Ga(4f) site, 
which sees 12 Cr neighbors, including three from the linking layer.~\cite{mendels}
Then in the pure system, Eq.~(\ref{eq-layermeans}) 
implies the mean susceptibility is exactly $\mu^2 /7J$ per spin, 
but the mean susceptibilty seen by Ga(4f) is $\mu^2 (1/4J-1/8J')$. 

\section {Why are the simplices satisfied?}
\label{sec-WhySat}

I now present three different -- not exclusive -- viewpoints 
for understanding simplex satisfaction.

\subsection{Single-impurity explanation}
\label{sec-singleimp}

The original explanation~\cite{Sh93}
just considered a single nonmagnetic impurity in a
pure background, as appropriate to the weak dilution regime. 
Ref.~\onlinecite{Sh93} exhibited a rearrangement
of a few spins (as few as 10) around this defect,  
as in Fig.~\ref{fig-defectloop},
which completely satisfies every simplex.  
Farther away, the spin deviations are strictly zero, in contrast to
an ordinary frustrated magnet (Sec.~\ref{sec-ordweak}). 
The general method -- first described for the pyrochlore case
~\cite{vill79} --
uses $q-1$ rotatable loops, each connecting a 
spin on one of the simplices containing the impurity, to a spin
on the other affected simplex. (``Rotatable''
means that, in the pure lattice, 
the spins on the loop can be rigidly rotated together to
produce other ground states.)
The deviations around the impurity
remind me of the screening around a test charge 
in a metal,  by the high density of excitations at zero energy --
in the present problem, those excitations are the rotatable loops.

Clearly this picture works for well-separated impurities --
but that requires $1-p$ to be quite small.
In Fig.~\ref{fig-defectloop}, all 19 spins must point in the 
pattern shown, modulo rotations; this conflicts with the 
pattern forced by a second impurity that sits anywhere on the two
hexagons in Fig.~\ref{fig-defectloop} or the eight hexagons
surrounding: a single impurity excludes other impurities on 40 other 
sites.
(The above counts and Fig.~\ref{fig-defectloop} assume a 
background consisting of the $\sqrt 3 \times \sqrt 3$ state;
in other coplanar  backgrounds, more spins are affected.)

The single-impurity picture, then,  breaks down at $p \approx 0.97$, 
where the defect configurations start to overlap.
To explain the simulations from this viewpoint, 
one is forced to postulate that the
defect spin deviations obey a nonlinear superposition principle,  
as magical as that of solitons in certain one-dimensional systems.

\subsection {Constraint propagation explanation}
\label{sec-constrprop}

This picture is most appropriate to the strong-dilution limit near $p_c$, 
where the connected cluster is tenuous.
Along a simple chain of sites, spins alternate, propagating order
as if there were a direct  bond between the endpoints. 
Now allow a few spins neighboring this path; these decorate the path 
with triangles which I will call ``ears'' (see Fig.~\ref{fig-earloops}).
Each ``ear'' removes constraints on the spins, 
since the two spins on the path are now constrained merely
to differ by a $120^\circ$ angle. 
When a path includes two or more ``ears'', 
there is no constraint at all between the endpoint spins.

As in Sec.~\ref{sec:ord-strong}, 
the key question is whether two paths which rejoin (forming a loop)
might propagate mutually exclusive constraints.
In contrast to the triangular case
of Fig.~\ref{fig-ordinary}(c)), 
{\it every simple loop has even length}. 
(Derived for the kagom\'e case in Ref.~\onlinecite{vd93}, Appendix;
for a general bisimplex lattice, it follows from the bipartiteness 
of the underlying network.)

All loops with the same number of ``ears'' are equivalent, 
since -- by the evenness lemma just stated --  they 
have the same number (modulo 2!) of ear-less links, 
at which the spins simply invert. 
Therefore it suffices to study loops of length 6
as in Fig.~\ref{fig-earloops}. 
The  one-eared loop shown 
{\it is} frustrated: 
the two spins in the ear should differ by $120^\circ$
due to the triangle, but by $180^\circ$ 
due to the five simple links connecting them.
On the other hand, the two-eared loop forces
the same $180^\circ$ angle as a simple chain (between the endpoint
spins where it connects to the rest of the world), and the
three-eared loop forces the same $120^\circ$ angles
as a giant triangle would.

The simulations described in Sec.~\ref{sec-simulations}
found 38 one-ear loops in 500 realizations of 100 cells
each at $p=0.55$, in full agreement with the predicted frequency
$p^7 [18 (1-p)^8 + 30(1-p)^9]$, per cell. 
Three other frustrated clusters were found, once each:
a one-eared 8-ring, and (in two variations) a
hexagon pair sharing two ears, with one added ear.
Each of these objects has (at most) 
one connection to the outside world 
forming the ``ear'', 
and thus has does {\it not} propagate frustration
on larger scales.

\subsection{Constraint counting}
\label{sec-constraints}

The preceding discussions
detected no frustration in either the strong or weak dilution regime, 
but are insufficiently general.
The constraint-counting (``Maxwellian'') approach of Moessner and
Chalker~\cite{moess98} is, I believe, the convincing
explanation of simplex satisfaction.
\footnote{
A caution, however, is that ``Maxwellian'' counting is a mean-field theory.
A threshold at which $D_g(p)$ vanishes is not generally the exact threshold, 
since portions of the structure may be under-constrained while others
are over-constrained. 
For related issues in elastic percolation, see
Ref.~\onlinecite{elasticperc}.}
The basic aim is to compute the dimensionality $D$ of the 
manifold of states in which all simplices are satisfied;
as long as this set is non-null (i.e. $D \ge 0$), it is obviously the 
ground state manifold $\XX$.

Now let $\tXXg$ be the manifold of ``generic'' simplex-satisfied states  
(having no linear relationships among the spins, 
apart from the $K$ constraints 
required to satisfy all the simplices).
The dimensionality of $\tXXg$ is
  \bee
       D_g =  F -K.
  \label{eq-constraints}
  \eee
Here $F$ is the number of degrees of freedom.
two per spin, so
$F/N = 2 p \nu$ per unit cell, where $\nusite$ is the number of
sites per unit cell,  as in Table~\ref{t-bisimplex}.
Table~\ref{t-constraints} gives the number 
of constraints per simplex, $\Ksim(q')$, where $q'$ is the number
of magnetic sites remaining after dilution.
Naively $\Ksim(q')=3$, for the three components of $\LS_\alpha=0$
(or its generalization  to nonzero magnetic field). 
However, in zero field when the simplex is a single bond ($q'=2$),  
the constraint is simply ``$\ss_2=-\ss_1$'': attaching $\ss_2$
does not change $D$, so the added constraints
must be two to cancel the added degrees of freedom.
Also, $\Ksim(0)= \Ksim(1)=0$. 
The total number of constraints is
   \bee
       K/N =  \sum _q \nu(q) C^{q}_{q'} p^{q'}(1-p)^{q-q'} \Ksim(q')
   \label{eq-Ksum}
   \eee
per unit cell, 
where $C^{q}_{q'}$ is the combinations of $q$ things taken $q'$ at a time, 
and $\nu(q)$ is the number of $q$-corner simplices per cell 
in the undiluted structure.  

Most relevant to propagating order (or frustration)
is the generic dimensionality 
$\Dgc$ of the extended connected cluster; 
I approximate this by subtracting from $D_g$
the contribution to $F$ by isolated sites.
The results are
   \bea
       \Dgc^{\rm kag}/N &=& 6p[1-(1-p)^4] -6p^2(2-p) , \nonumber \\
       \Dgc^{\rm sand}/N &=& p[14-12(1-p)^5-2(1-p)^6]
                     - 12p^2(p^3 -3p +3)  , \nonumber \\
       \Dgc^{\rm pyr}/N &=& 8p[1-(1-p)^6] -12 p^2(1-p)(2-p) , 
   \label{eq-Dgcp}
   \eea
for the three lattices. 
Counterintuitively, removing  a single site 
from the kagom\'e lattice leaves $D_g$ unchanged, but
removing an adjacent pair of sites increases $D_g$ by 1.
In a general bisimplex lattice, 
after dilution to
the point $\ptri$) at which the average 
simplex is a triangle, 
further dilution ought to (slightly) increase $\Dgc$.
(Note $\ptri \approx 1$, $0.9$, and $0.75$ for the kagom\'e, 
sandwich, and pyrochlore lattices respectively.)
Indeed, $\Dgc(p)$ at first decreases rapidly with dilution and
and tends to level off below $\ptri$, at roughly 
$\Dgc(p) \sim 0.2$ on the kagom\'e, 
$\sim 1.6$ on the sandwich (SCGO) lattice, 
and $\sim 1$ on the pyrochlore, per unit cell.
Since $\Dgc(p)$ remains positive,  we expect simplex-satisfied ground
states at any $p>p_c$.

\subsubsection{Generic/non-generic state crossover?}

Do not be misled by the term ``generic''!
The typical (physical) ground state is ``non-generic'' if such 
states have a higher dimensionality than the ``generic'' ones.
In particular, in the {\it pure} kagom\'e lattice, starting with 
the $\sqrt{3}\times \sqrt{3}$ coplanar state, one 
can rotate one of every three six-loops (``weathervanes'')
by an independent angle. 
This ground-state manifold has
dimensionality $D_n= 1/3$ per cell which dominates
$D_g = 0$.  Such states are non-generic because some spins (e.g. 
every second one in a rotatable loop) have {\it exactly} the same directions.
(Indeed, in ground states relaxed from a random
spin configuration, second-neighbor spins often point
in nearly the same direction.)  
For small dilution, the non-generic states still dominate;
these are precisely the coplanar states with rare impurities
discussed in Sec.~\ref{sec-singleimp}.  But the spin rearrangement
at each impurity (Fig.~\ref{fig-defectloop}) touches, 
and thus immobilizes,  10 six-loops, and the 
frequency of impurities is $3(1-p)$ per cell.
So, I estimate
   \bee
       D_n (p) /N  \approx \frac{1}{3} (1-30[1-p]) 
   \eee
as the dimension per unit cell of the non-generic manifold.
The non-generic manifold loses out to the generic one
at $p\approx 0.97$, in the kagom\'e lattice. 
By contrast, the pure pyrochlore
lattice has $D_g(1)=2$,  while it appears $D_n(1)=1$
(and presumably $D_n(p)$ also plummets upon dilution). 
Thus, only the generic manifold is relevant in the pyrochlore. 
(essentially this paraphrases the absence
of (local) ``order-by-disorder''~\cite{moess98}.)

\subsubsection {Back to frustration}
\label{sec-frust}

Up to now, I considered the
isotropic classical Heisenberg  antiferromagnet, with possible
dilution.  
The degeneracies retained by the classical ground state manifold, 
even under dilution, may be broken by
various realistic perturbations.
These can be handled by the constraint-counting framework,  
as in Table~\ref{t-constraints}
(with extensions to accomodate constraints
that involve two simplices).

Bond disorder has been identified
in pyrochlore antiferromagnets~\cite{pyro-dis,mire00HFM} 
and was modeled theoretically~\cite{hold00HFM}.
It can produce ``canted local states''~\cite{vill79}
of deviated spins in ordinary Heisenberg antiferromagnets 
(interacting as a spin glass~\cite{vill79}).
In a bisimplex lattice.  it has a different effect: 
the constraints $\Ksim(q)$ are greatly increased.
Assuming a fraction $x_F$ of ferromagnetic bonds, 
inserting $\Ksim(q)$ from Table~\ref{t-constraints} 
into Eq.~(\ref{eq-constraints}) gives
   \bea
          D_g^{\rm pyr}  & =& 2(1-18 x_F), \nonumber \\
          D_g^{\rm sand.} &=&   2(1-21 x_F) 
   \label{eq-DgFM}
   \eea
in place of (\ref{eq-Dgcp}). 
Eq.~(\ref{eq-DgFM}) predicts that 
$x_F\approx 0.05$ is a threshold, 
beyond which the system becomes overconstrained and (presumably)
spin-glassy.

Quantum fluctuations also impose collinearity in a tetrahedron.
For easy-{\it axis} exchange anisotropy, 
rather surprisingly the $q=3$ ground state manifold's
dimensionality is unchanged~\cite{miya85}.
(The constraint count for easy-{\it plane} exchange anisotropy 
is the same as for XY spins, see ~\onlinecite{moess98}). 
Magnetic field is a special case (already covered
in Sec.~\ref{sec-orphan}).  The spins along a 
chain when $\BB\neq 0$ alternate between two possible directions, 
indicating that each extra link contributes only two constraints, 
as accounted in footnote $c$ of Table~\ref{t-constraints}. 
Substitution into (\ref{eq-Ksum}) and (\ref{eq-constraints})
shows that most of these perturbations make the
generic states overconstrained,  and presumably glassy, 
though (in the undiluted cases) one 
needs to rule out possible non-generic ground states. 

\section{Conclusions}

Large-$S$, non-random antiferromagnet should have periodic
long-range order at $T=0$, even on a highly frustrated lattice.
A purely classical picture is invalid, even for $S \gg 1$, 
when temperature is far below the spin-wave energy $jS$, 
but this may be fixed up by using the effective Hamiltonians
in the simulation (Sec.~\ref{sec-pitfalls}). 

Dilution does {\it not} engender a spin-glass in classical ground states,
as it does in ordinary frustrated magnets, since simplices remain
satisfied, as confirmed by simulations (Sec.~\ref{sec-simulations}). 
This was understood most generally from constraint-counting arguments
(Sec.~\ref{sec-constraints}.)
The observed spin-glass state in SCGO {\it might} be
attributed to a competition of dilution with the coplanarity
tendency due to quantum fluctuations. 
Effective Hamiltonians (Sec.~\ref{sec-Heff}), 
serve as a first (perhaps only) way to model the 
zero-point energy contribution in a disordered lattice
-- even if the functional form is not quite right --
since a more exact treatment would be intractable, 
in the presence of the multiple perturbations
of Sec.~\ref{sec-frust}.

Some physical corollaries are deduced about the structure factor, 
paramagnetic susceptibility, and NMR response, including
a ``Gauss's Law'' rule (Sec.~\ref{sec-gauss})
which manifests the long-range effects of the
paramagnetic (``half-orphan'') spins.
Experiments sensitive to percolation on SCGO ought
to be performed with occupations below the threshold
$p_c \approx 0.5$, which is estimated here for the first time
(Appendix~\ref{app-pc}). 

I speculated that 
-- in the plain kagom\'e lattice, at least --
there is a threshold $\pstar \approx 0.98$
separating two regimes of dilution: nongeneric coplanar states, 
peppered with defects, at $p > \pstar$, but at $p <  \pstar$ 
generic states, in which every trace is lost of the coplanar background
(Secs.~\ref{sec-singleimp} and \ref{sec-constraints}). 
This would imply, of course, that it is invalid to extrapolate
experimental measurements for $p \in (0.9, 0.95)$ up to $p=1$.~\footnote{
There is less likely to be a $\pstar$ in the sandwich (SCGO)
lattice case, but I do not understand the non-generic
states of that system well enough to guarantee  this.}

\acknowledgements

I thank E. F. Shender, 
V. B. Cherepanov, C. Broholm. 
P.~Mendels, and I.~Mirebeau
for stimulating discussions, 
and I thank Johns Hopkins University for
hospitality (in 1994) when parts of this
work were begun.
This work was supported by NSF Grant No. DMR-9981744, and
used the Cornell Center for Materials Research 
computing facilities, 
supported by NSF MRSEC DMR-9632275.

\appendix
\section {Percolation threshold of kagom\'e sandwich lattice}
\label{app-pc}

A very simple program (no spins are involved) generated
random configurations with a fixed number of diluted sites.
Having found the number of sites $m_j$ in the $j$-th connected cluster,
and letting $m_1$ be the largest of them, 
I computed two percolation quantities:
the ``percolation susceptibility'' $\chi_p = N^{-1} \sum_{j \neq 1} m_j^2$,
and $P_p = m_1/pN $, 
the fraction in the extensive (``infinite'') cluster.
In the limit $N\to \infty$, 
one expects $\chi_p \sim |p-p_c|^{-\gamma}$ on either side of
$p_c$, while $P_p \sim |p-p_c|^\beta$ on the high side, 
and $P_p\equiv 0$ on the low side.

From the behavior of these two quantities, I
estimated $p_c \approx 0.50(2)$.  
To probe the systematic error due to size-dependence, 
several sizes were tried
(up to $16\times 16$, i.e. 1792 sites before dilution);
however, a genuine scaling fit was not carried out.

\end{multicols}

\newpage

\begin{table}
\caption{Bisimplex lattices. Number of sites $\nusite$ and of 
simplices $\nu(q)$ are per Bravais unit cell.}
\label{t-bisimplex}
\begin{tabular}{|l|l|l|c|c|c|c|c|c|} 
Name &  Derived from & Bravais lattice & $d$ & $(q,q')$ &  
        $\nusite$ & $\nu(3)$ & $\nu(4)$& $p_c$ \\ \hline
kagom\'e &   honeycomb & triangular &  2  & 3,3 & 3 &2& 0& $0.65$ \\
garnet   & ``gyroid'' graph & bcc   &  3  & 3,3 & 6 &4& 0& $>$ 0.5 ? \\
crossed-square & square & square    &  2  & 4,4 & 2 &0& 1& $0.50$  \\
pyrochlore &  diamond  &  fcc       &  3  & 4,4 & 4 &0& 2& $0.39$ \\
kag. sandwich &  --    & triangular &  2  & 3,4 & 7 &2& 2& $0.50(2)$ \\
\end{tabular}
\end{table}

\begin{table}
\caption{Constraint count $\Ksim(q)$ in $q$-corner simplices} 
\label{t-constraints}
\begin{tabular}{|l|c|c|c|} 
Case &  $q=2$&   $q=3$ &  $q=4$
           \\ \hline
Isotropic Heisenberg spins\tablenote{
    Also the case $J'\neq J$}
                           &       2  &   3&   3  \\
The same, with spin waves     &       2  &   3\tablenote{
  For the coplanarity constraint, add $+1$ for every site
  with $q=3$ on both sides,}                
                                             & 5 \\
One ferromagnetic bond     &       2  &   4  & 6 \\
Easy-axis exchange anisotropy &    4 &    4 &  5 \\
Magnetic field\tablenote{
Also $\Ksim(1)=2$, since $\ss_i \parallel \BB$
in that case.}                &     3 \tablenote{
     But subtract $1$  for every site with $q=2$ on both sides of it.}
                                     &    3 &  3 \\
\end{tabular}
\end{table}

\begin{centering}

\begin{figure}
\begin{minipage}{13cm}
\epsfxsize=13cm \epsfbox{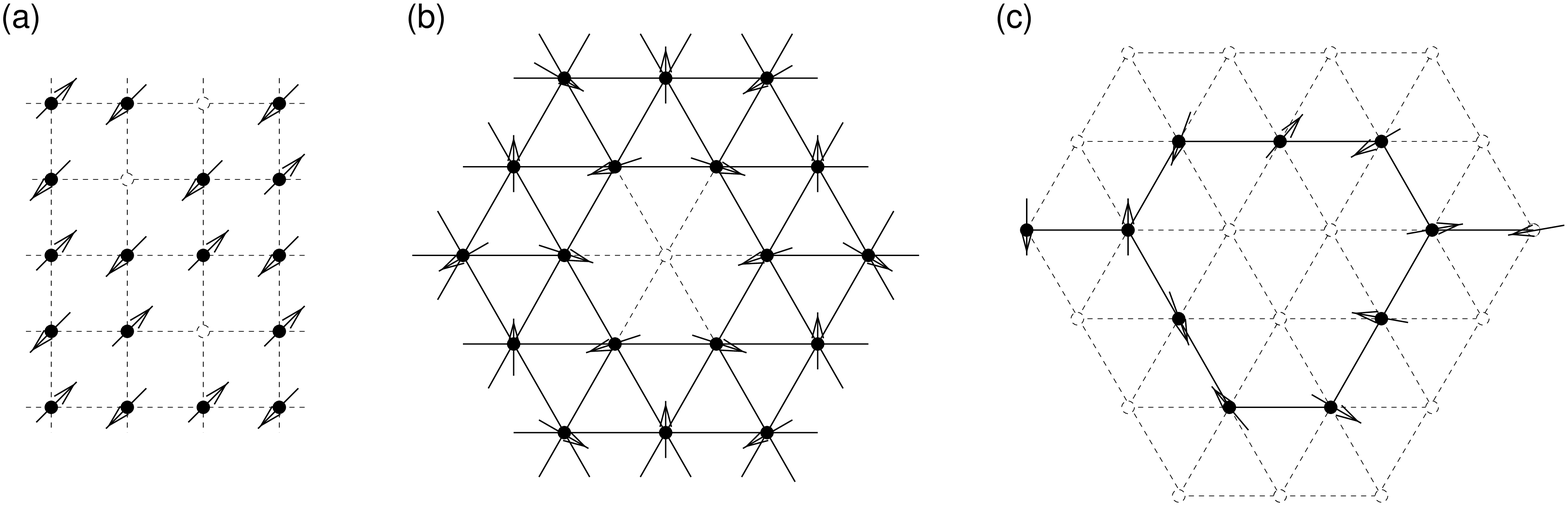}
\caption{
Diluting ordinary antiferromagnets;
diluted sites and their bonds are always shown dashed.
(a). Unfrustrated antiferromagnet:
the total moment here is one down spin, the excess of odd sites over even sites
(b). Removal of one spin in the triangular Heisenberg
antiferromagnet, causing neighboring spins
to deviate in the direction of the missing spin.
Spins on the second ring outwards are given the directions they
would have in the pure lattice, though in reality a small distortion
is found at any radius. 
(c). A loop of sites with an odd number of steps 
introduces random frustration
and spin relationships absent in the pure lattice. 
The upper (lower) path favors the endpoint spins to be parallel
(antiparallel). }
\label{fig-ordinary}
\end{minipage}
\end{figure}

\begin{figure}
\begin{minipage}{13cm}
\epsfxsize=6 cm \epsfbox{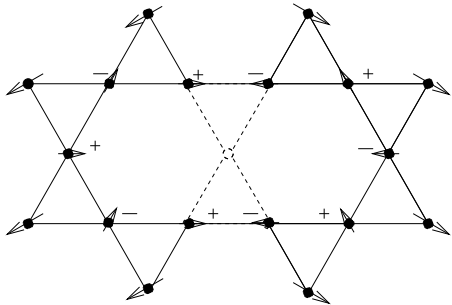}
\caption{
One removed spin in the kagom\'e antiferromagnet, and the
10-site defect loop induced around it.
Circles with dots or crosses are spins pointing directly
out of or into the paper; ``$+$'' and ``$-$'' signs 
indicate spins with an outward or inward component.
The surroundings are part of a coplanar $\sqrt{3}\times\sqrt{3}$
ground state.}
\label{fig-defectloop}
\end{minipage}
\end{figure}

\begin{figure}
\begin{minipage}{13cm}
\epsfxsize=13cm \epsfbox{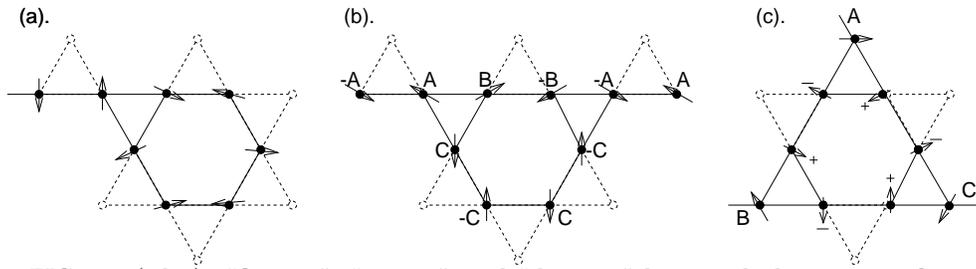}
\caption{
(a,b,c).  ``One-ear'',  ``two-ear'',  and ``three-ear'' 
loops with their spin configurations (unique, modulo rotations in spin space). 
``A'', ``B'', and ``C'' mark three spin directions which differ
by $120^\circ$.}
\label{fig-earloops}
\end{minipage}
\end{figure}

\end{centering}

\end{document}